\documentclass[aps,prx,twocolumn,superscriptaddress,showpacs]{revtex4-1}
\usepackage{hyperref}
\usepackage{times,xspace}
\usepackage{graphicx,graphics,color,epsfig}
\usepackage{amsmath}
\usepackage{amssymb}
\usepackage{amsfonts}
\usepackage{amsfonts}
\usepackage{epstopdf}
\usepackage{bm}
\usepackage{hyperref}
\usepackage{longtable}     
\usepackage{url}           
\usepackage{bm}            
\usepackage{color}
\usepackage{float}
\usepackage[normalem]{ulem}

\newcommand{\blue}{\textcolor{blue}}

\begin{document}
		
	\title{Prediction of $f$-wave pairing symmetry in YBa$_2$Cu$_3$O$_{6+x}$ cuprates}
	
	\author{Priyo Adhikary, and Tanmoy Das}
	\affiliation{Department of Physics, Indian Institute of Science, Bangalore, Karnataka 560012, India}
	\date{\today} 

\begin{abstract}
We perform a numerical simulation of a three-band Hubbard model with two CuO$_2$ planes and a single CuO chain layer for YBCO cuprates. The spin-fluctuation mediated pairing interaction is computed within the multiband random-phase approximation, and its pairing eigenvalues and eigenfunctions are solved as a function of chain state filling factor $n_c$. We find that for the intrinsic value of $n_c$ in YBCO samples, one obtains the usual $d$-wave pairing symmetry. However, if we dope the chain layers with holes, while keeping the plane states doping fixed, the leading pairing symmetry solution becomes an unconventional $f$-wave symmetry. The mechanism behind the $f$-wave pairing is the competition between the plane states antiferromagnetic nesting and chain states' uniaxial nesting. We also find that the pairing strength is strongly augmented when the flat band bottom of the chain state passes the Fermi level for a fixed plane states doping. The $f$-wave pairing symmetry can be realized in YBCO cuprates in future experiments where the self-doping mechanism between the chain and plane states can be minimized so that only chain state can be selectively hole doped. 
\end{abstract}


\maketitle

\section{Introduction}\label{theory}
In cuprate superconductors, $d$-wave pairing symmetry is well established in all member materials at most of the doping ranges.\cite{TsueiRMP,ScalapinoDwave,KirtleyDwave,SCcuprates} Supporting evidence to the $d$-wave pairing symmetry come from various complementary studies including junction experiments,\cite{JunctionExp} spectroscopic fingerprints of the nodal pairing states,\cite{ARPES,VishikReview,STM} as well as power-law dependence in various thermodynamical and transport measurements.\cite{KirtleyDwave,dwavePDepth,dwaveNMR,dwaveOptics,dwaveCv,dwaveBAngle,ScalapinoDwave,Taillefer} There have been few but robust contradictory evidence to the nodal superconducting (SC) gap in a limited doping region in several cupates. Notably, in electron-doped cuprates, in the deep underdoped region, various measurements exhibited the presence of nodeless SC gap, which was initially assumed to be an $s$-wave pairing symmetry.\cite{wu,andreone,schneider,alff,skinta,snezhko,biswas,qazilbash,chesca,ariando} Later on, it was shown that the underlying pairing state has the $d$-wave symmetry, however owing to the loss of Fermi surface (FS) at the nodal region due to antiferromagnetic order, the effective quasiparticle spectrum looses its gapless features.\cite{TanmoyED,CSTing} Furthermore, more recently, there has been convincing evidence of nodeless SC gap in the deep underdoped region of La-based,\cite{LSCO2000,LSCO2013} Bi-based,\cite{bi22122006,bi2212,bi2201} Cl-based,\cite{ccoc} and Yb-based hole-doped cuprates.\cite{YBCO} Theoretical explanation to this mechanism is still divided into whether an underlying $d$-wave state looses its nodal state due to correlation\cite{CDMFT1,CDMFT2} or disorder,\cite{CoulombGap} or a new pairing state arises here.\cite{Annica,TSC,DasNodelessSC,Gupta} However, so far there has not been any experimental indication or theoretical prediction for an $f$-wave pairing symmetry in cuprates. 

Our present work focuses on YBa$_2$Cu$_3$O$_{6+x}$ (YBCO$_{6+x}$) systems. YBCO lattice structure is special compared to other cuprates. Here the lattice comprises an alternate stacking of two CuO$_2$ square blocks within the $ab$-plane, and a CuO chain layer oriented along the $b$- direction. We, henceforth, denote the corresponding states as plane and chain states, respectively. Oxygen doping introduces holes on the CuO$_2$ plane states, and YBCO$_6$ and YBCO$_7$ compounds represent undoped and overdoped samples, respectively, while superconductivity arises in between these two compositions. Prior density-function theory (DFT) calculations\cite{DFTYBCO} showed that the chain state is absent from the Fermi level in the undoped (YBCO$_6$) compound, while it crosses the Fermi level for the finite doping region. Photoemission measurement also exhibited the evidence of quasi-1D chain states on the Fermi level.\cite{damascelli2008,damascelli2010,ARPES_borisenko} Various transport measurements consistently pointed out that the chain states are highly metallic.\cite{Hussey,Riggs} Moreover, at finite dopings, the chain state strongly hybridizes with the plane states near the magnetic zone boundary, establishing that the electron tunneling and/or charge transfer between the chain and plane states are strong enough to play an important role on the low-energy properties of YBCO cuprates.\cite{Hussey,Riggs,Xiao,Magnuson,Uimin,TcHighYBCO,Combescot,YBCOdensitymod,Dasnematic,DasQOchain,atkinson}

In this work we study how the SC pairing symmetry and pairing strength are modified when the contributions of the chain states are included in the calculations. We consider a three-band tight-binding model with two planes and one chain state per unit cell.\cite{Dasnematic,DasQOchain,atkinson} We construct the pairing potential arising from the spin-fluctuation mechanism; the many-body interaction is captured with the multiband Hubbard model within the weak-coupling random-phase  approximation (RPA).\cite{SCrepulsive,SCcuprates,SCpnictides,SCHF,SCorganics,SCTMDC,SCGraphene} The leading eigenvalue and its corresponding eigenfunction of the static pairing potential gives the SC coupling constant and the pairing symmetry of the system, respectively. The basic understanding of the spin-fluctuation mediated pairing symmetry is that when the FS nesting is strong at a preferential wavevector, say ${\bf Q}$, it leads to a pairing symmetry which changes sign across the momentum ${\bf k}$ and ${\bf k}+{\bf Q}$ on the FS.\cite{SCrepulsive,SCcuprates,SCpnictides,SCHF,SCorganics,SCTMDC,SCGraphene}In cuprates, the FS nesting is dominated by the spin-fluctuation wavevector ${\bf Q}=(\pi,\pi)$ which connects the Fermi momenta near the `magnetic hot-spot' (MHS) (where the plane FS meets the magnetic zone boundary), and one obtains a $d_{x^2-y^2}$-wave solution.\cite{SCcuprates}. 

\begin{figure}
\includegraphics[width=0.5\textwidth]{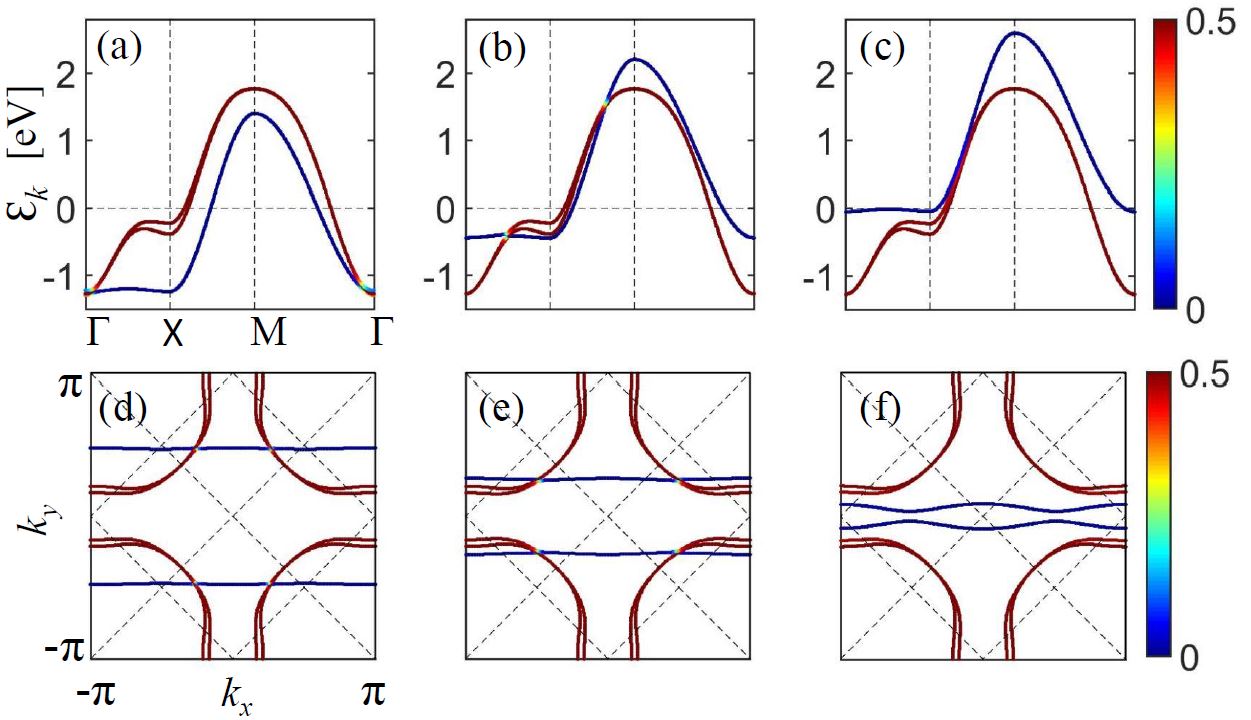}
\caption{(a-c) Electronic structures of the three-band noninteracting model, Eq.~\eqref{Ham} at three representative doping values on the chain states, while the doping on the plane state is kept fixed. (d)-(f) Corresponding FSs are shown for the same three cases presented in the upper panel. Red to blue color map in a given band at a ${\bf k}$- point gives the orbital contribution from the plane and chain states, respectively. (a), (d) When the chain state is highly electron doped, the HHS lies above the diagonal direction of the Brillouin zone, where an $f$-wave pairing symmetry is obtained. (b), (e) At the intermediate electron-doping on the chain state, which is realized in single crystal YBCO samples, the HHS moves below the BZ diagonal direction, and here we obtain $d$-wave pairing solution. (c), (f) A characteristic doping where the bottom of the chain band just lies at the Fermi level, giving high-density of states at the Fermi level, and hence SC strength reaches its optimum value as a function of chain state doping for a fixed plane doping.} 
\label{fig:orb_wgt} 
\end{figure}

Recent experimental studies have achieved selectively doping only the chain state, while the plane state maintains nearly a fixed doping level.\cite{Magnuson,YBCOdensitymod,Middey} Motivated by this, we consider the doping variation of the chain state for various fixed doping concentrations on the plane state across its optimal doping regime. We find that for the natural doping ranges of the chain state, the pairing symmetry is $d_{x^2-y^2}$-wave. {\it But as the chain doping is tuned above some critical value, which is not naturally achieved in YBCO$_{6+x}$ single crystals, the pairing symmetry on the plane states is changed to an $f$-wave pairing symmetry.} We find that this pairing symmetry transition is linked to where the plane and chain states are hybridized in the Brillouin zone (BZ). Let us call the momentum point where the chain and plane states' FSs meet as `hybridization hot-spot' (HHS), see Fig.~\ref{fig:orb_wgt}. We find that when the HHS lies below the nodal line (diagonal direction of the BZ), the pairing symmetry is $d$-wave like, see Figs.~\ref{fig:orb_wgt}(b) and ~\ref{fig:orb_wgt}(c). The pairing symmetry changes to an $f$-wave symmetry when the HHS crosses above the BZ diagonal directions, i.e., when the chain state is highly electron-doped, see Fig.~\ref{fig:orb_wgt}(a). This conclusion is found to be robust for a wide range of interaction strength as well as for various values of the hybridization strength between the two layers. The $f$-wave pairing symmetry has not yet been reported in YBCO$_{6+x}$ samples, but with the advent of layered dependent doping mechanism, such a pairing symmetry can be achieved in future experiments with electron doping on the chain states. 

The rest of the paper is arranged as follows. In Sec. \ref{Model}, we discuss our model. This section includes discussions on the tight-binding model, susceptibility calculation details, and the derivation of the density-fluctuation mediated pairing interaction. In Sec. \ref{results}, we present our results of FS topologies, corresponding FS nesting profiles, and pairing symmetries at several representative chain state's dopings. We also present the results of pairing strength and pairing symmetry for a large plane and chain doping ranges. Finally, we discuss the robustness of the results with various plane-chain hybridization, and interaction strengths. We discuss and conclude our results in Sec.~\ref{discuss}.

\section{Model}\label{Model}

\subsection{Tight binding model}\label{tbham}
We consider a three band model in which two CuO$_2$ layers are interacting with an uniaxial CuO chain state.\cite{Dasnematic,DasQOchain,atkinson} We work in the basis of $\Psi_{\sigma}({\bf k})$=($c_{p\sigma}({\bf k})$, $c_{p'\sigma}({\bf k})$, $c_{c\sigma}({\bf k})$)$^T$, where $c_{\alpha\sigma}({\bf k})$ annihilates an electron on the $\alpha^{\rm th}$ layer with momentum ${\bf k}$, and spin $\sigma=\uparrow/\downarrow$, and the subscript $\alpha=p,p'$ refers to the two planes, and $\alpha=c$ stands for the chain layer. In this spinor, the Hamiltonian reads as:
\begin{eqnarray}
H = \left(\begin{matrix}
\xi_{p}    & \xi_{pp'} & \xi_{cp}  \\
\xi^*_{pp'}   & \xi_{p'}  & \xi_{cp'} \\
\xi^*_{cp}   &\xi^*_{cp'}  & \xi_{c} 
\end{matrix}\right).
\label{Ham}
\end{eqnarray}
({\bf k} dependence in all terms above are suppressed for simplicity). Here $\xi_{p/p'}$, and $\xi_{c}$ are the intralayer dispersions within the plane and chain states, respectively. $\xi_{pp'}$ and $\xi_{cp}$ are the interlayer hoppings between the two planes and between plane and chain states, respectively. The corresponding dispersion terms are obtained within the tight-binding model including nearest and various next-nearest neighbor hoppings as appropriate to describe the corresponding DFT band structure (see Refs.~\cite{Dasnematic,DasQOchain}). Following the DFT result of a weak $k_z$ dispersion in this compound,\cite{DFTYBCO} we neglect three dimensional dispersion. The explicit form of the dispersions are 
\begin{subequations}
\begin{eqnarray}\label{intraband}
\xi_{p}&=&-2t(c_{x}+ c_{{y}})+2t^\prime c_x c_{y}+2t^{\prime\prime}\left(c_{2x}+ c_{2y}\right) - \mu_{p},\\
\xi_{c}&=& -2t_{cy}c_{y}-2t_{cx}c_{2x} -  \mu_{c},\\
\xi_{pp'}&=&-2t_{pp}\left(c_x- c_{y}\right)^{2},\\
\xi_{cp}&=&t_{cp}.
\end{eqnarray}
\end{subequations}
$\mu_{p,c}$ are the onsite potentials for the plane and chain states. We use the brief notation of $c_{i\alpha}=\cos{(i\alpha)}$, where $i$ dictates the interatomic distances in units of lattice vectors, and $\alpha=k_{x,y}$. We obtain the tight-binding parameters by fitting to the DFT band structure:$(t, t^\prime, t^{\prime\prime},t_{cy},t_{cx}, t_{pp}, t_{cp}, \mu_{p}, \mu_{c})=(0.38, -0.18, 0.25, 0.66, 0.01, -0.01 , 0.02, -0.37, -1.15)$ eV. We consider the anisotropy along the \textbf{a} axis for the chain band by setting t$_{cx}<<$ t$_{cy}$, giving the chain band to be very much uniaxial along the $\textbf{b}$ axis. 

We diagonalize the Hamiltonian in Eq. (~\ref{Ham}) and obtain three eigenvalues $E_{\nu}({\bf k})$ and corresponding eigenvectors $\phi^{\nu}_{\alpha}({\bf k})$, where $\nu$ denotes band indices, and $\alpha$ stands for layer species. We assume the operator for annihilating a quasiparticle in the $\nu^{\rm th}$-band with spin $\sigma$ is $\gamma_{\nu\sigma}({\bf k})$. Then the spinor in the eigenbasis is $\Phi_{\sigma}({\bf k})$=($\gamma_{1\sigma}({\bf k})$, $\gamma_{2\sigma}({\bf k})$, $\gamma_{3\sigma}({\bf k})$)$^T$. 

The density operators for the $i^{\rm th}$ layer for the spin $\sigma$ is ${\rm n}_{i\sigma}({\bf q})=\frac{1}{\Omega_{\rm BZ}}\sum_{\bf k}c_{i{\bf k}\sigma}^{\dag}c_{i{\bf k}+{\bf q},\sigma}$. We fix the charge density for plane and chain states separately by self-consistently evaluating the density operators at ${\bf q}\rightarrow 0$. The electron concentration on the plane state is taken as average over the two planes $n_p = 2\frac{1}{2}(\langle {\rm n}_p\rangle + \langle {\rm n}_{p'}\rangle)$, and that for the chain state is $n_c = 2\langle {\rm n}_c\rangle$. Here the factor 2 originates from spin-degeneracy. The thermal average is taken over {\it all} eigenstates with $\langle \gamma_{\nu\sigma}({\bf k})\rangle = f(E_{\nu}({\bf k}))$ as the Fermi Dirac distribution function. We self-consistently fix the value of $n_p$ and $n_c$ by treating  $\mu_p$ and $\mu_p$ as free parameters.

\subsection{Multiband RPA susceptibility}\label{rpasus}

Next, to study the modulation of FS nesting profile and feed the corresponding information to the spin-fluctuation mediated pairing potential, we consider a multiband Hubbard model:
\begin{eqnarray} 
H_{\rm int} &=& \sum_{\alpha\in p,p',c}U_{\alpha} {\rm n}_{\alpha\uparrow}{\rm n}_{\alpha\downarrow}+ \sum_{\substack{\alpha\ne \beta\in (p,p',c) \\ \sigma\sigma'\in(\uparrow,\downarrow)}}V_{\alpha \beta}{\rm n}_{\alpha\sigma}{\rm n}_{\beta\sigma'}.
\label{Hint}
\end{eqnarray}
$U_p=U_{p'}$ is the onsite Hubbard interaction within the two plane layers, while $U_c$ is the same for the intrachain layer. $V_{p}$, $V_{c}$ are the onsite Hubbard interaction between the two planes, and plane-chain layers, respectively. Hund's coupling between these layers (all with $d_{x^2-y^2}$ orbitals symmetry) is ignored. By expanding the interaction term to multiple interaction channels, and collecting the terms which give a pairing interaction (both singlet and triplet channels are considered), we obtain the effective pairing potential $\Gamma_{\alpha\beta}^{\gamma\delta}({\bf q})$ as\cite{SCrepulsive,SCcuprates,SCpnictides,SCHF,SCorganics,SCTMDC,SCGraphene}
\begin{eqnarray}
H_{\rm int} &\approx& \frac{1}{\Omega_{\rm BZ}^2}\sum_{\alpha\beta\gamma\delta}\sum_{{\bf kq},\sigma\sigma'} \Gamma_{\alpha\beta}^{\gamma\delta}({\bf q})\nonumber\\
&&\quad\times c_{\alpha \sigma}^{\dagger}({\bf k})c_{\beta\sigma'}^{\dagger}(-{\bf  k})c_{\gamma\sigma'}({\bf -k-q})c_{\delta\sigma}({\bf k+q}).
\label{Hintpair}
\end{eqnarray}
$\sigma'=\pm\sigma$ give triplet and singlet pairing channels, respectively. This pairing potential, obtained in Refs.~\cite{SCrepulsive}, includes a summation of bubble and ladder diagrams within the random phase approximation (RPA). The pairing potential in general involves four orbital indices and thus is a tensor in the orbital basis. We denote all such tensors by the `tilde' symbol. The pairing potentials in the singlet ($\tilde{\Gamma}_{\uparrow\downarrow}$) and triplet ($\tilde{\Gamma}_{\uparrow\uparrow}$) channels are
\begin{subequations}
\begin{eqnarray}
\tilde{\Gamma}_{\uparrow\downarrow}({\bf q})&=&\frac{1}{2}\big[3{\tilde U}_{s}{\tilde \chi}_{s}({\bf q}){\tilde U}_{s} - {\tilde U}_{c}{\tilde \chi}_{c}({\bf q}){\tilde U}_{c} + {\tilde U}_{s}+{\tilde U}_{c}\big],
\label{singlet}\\
\tilde{\Gamma}_{\uparrow\uparrow}({\bf q})&=& -\frac{1}{2}\big[{\tilde U}_{s}{\tilde \chi}_{s}({\bf q}){\tilde U}_{s} + {\tilde U}_{c}{\tilde \chi}_{c}({\bf q}){\tilde U}_{c}- {\tilde U}^{s}-{\tilde U}_{c}\big].
\label{triplet}
\end{eqnarray}
\end{subequations}
Here subscript `s' and `c' denote spin and charge fluctuation channels, respectively. $\tilde{U}_{s/c}$ are the onsite interaction tensors for spin and charge fluctuations, respectively, defined in the same basis as $\tilde{\Gamma}$. Its nonvanishing components are $(\tilde{U}_{s,c})_{\alpha\alpha}^{\alpha\alpha}=U_{p/c}$ for intraplane ($\alpha={\rm p,p'}$) and intrachain ($\alpha={\rm c}$) layers. According to the definition in Eq.~\eqref{Hint}, the inter-plane Coulomb interaction enters into $(\tilde{U}_{s,c})_{pp}^{p'p'}=V_{p}$, and plane-chain interaction is $(\tilde{U}_{s,c})_{pp}^{cc}=(\tilde{U}_{s,c})_{p'p'}^{cc}=V_{c}$.

$\tilde{\chi}_{s/c}$ are the density-density correlators (tensors in the same orbital basis) for the spin and charge density channels. We define the noninteracting density-density correlation function (Lindhard susceptibility) $\tilde{\chi}_{0}$ within the standard linear response theory:\cite{SCTMDC}  
\begin{eqnarray}
[\chi_{0}(\textbf{q})]_{\alpha\beta}^{\gamma\delta}&=&-\frac{1}{\Omega_{\rm BZ}}\sum_{{\bf k},\nu\nu'}\phi^{\nu}_{\beta}({\bf k})\phi^{\nu\dagger}_{\alpha}({\bf k})\phi^{\nu'}_{\delta}({\bf k+q})\phi^{\nu'\dagger}_{\gamma} ({\bf k+q})\nonumber\\
&&\qquad~~~\times \frac{f(E_{\nu'}({{\bf k+q}}))- f(E_{\nu}({{\bf k}}))}{E_{\nu'}({\bf k+q})-E_{\nu}({\bf k})+i\epsilon}.
\label{Lindhard}
\end{eqnarray}
Many body effect of Coulomb interaction in the density-density correlation is captured within $S$-matrix expansion of Hubbard Hamiltonian in Eq.~\eqref{Hint}. By summing over different bubble and ladder diagrams we obtain the RPA spin and charge susceptibilities as: 
\begin{eqnarray}\label{spin_sus}
\tilde{\chi}_{\rm s/c}({\bf q})= \tilde{\chi}_{0}({\bf q})\left(\tilde{\mathbb{I}} \mp \tilde{U}_{s/c}\tilde{\chi}_{0}({\bf q})\right)^{-1},
\label{RPA}
\end{eqnarray}
where $\tilde{\mathbb{I}}$ is the unit matrix. We notice that the strong FS nesting features captured within the Lindhard susceptibility in Eq.~\eqref{Lindhard} are automatically translated to strong peaks in the RPA susceptibilities in Eq.~\eqref{RPA}. The RPA denominator of the spin susceptibility, having value $<1$, enhances the FS nesting strength in the bare susceptibility $\tilde{\chi}_{0}({\bf q})$. On the other hand, the RPA denominator for the charge channel is $>1$ suppressing the charge fluctuations. In addition, the zeros of the RPA denominator in the spin channel gives gapless magnon modes. The amplitude of the magnon modes are strongly suppressed in the optimal hole doping region of YBCO, being away from the AFM critical point.\cite{DasPanagopoulos,vishik2017,KivelsonNature}. Finally, all the strong FS nesting features in the RPA susceptibilities directly enter into the SC pairing channels through Eqs.~\eqref{singlet}, and \eqref{triplet} and determine the pairing symmetry accordingly.

\subsection{Superconducting pairing symmetry}\label{scpairsym}
Equation ~\eqref{Hintpair} gives the pairing interaction for pairing between orbitals. However, we solve the BCS gap equation in the band basis. To make this transformation, we make use of the unitary transformation $c_{\alpha\sigma}\rightarrow \sum_{\nu}\mathcal{U}^{\alpha}_{\nu}\gamma_{\nu\sigma}$ for all ${\bf k}$ and spin $\sigma$. With this substitution we obtain the pairing interaction Hamiltonian in the band basis as
\begin{eqnarray}
H_{\rm int} &\approx& \sum_{\nu\nu'}\sum_{{\bf kq},\sigma\sigma'} \Gamma'_{\nu\nu'}({\bf k,q})\nonumber\\
&&~\times \frac{1}{\Omega^2_{\rm BZ}}\gamma_{\nu \sigma}^{\dagger}({\bf k})\gamma_{\nu\sigma'}^{\dagger}(-{\bf k})\gamma_{\nu'\sigma'}({\bf -k-q})\gamma_{\nu'\sigma}({\bf k+q}).
\label{Hintpairband}
\end{eqnarray}
The same equation holds for both singlet and triplet pairing and thus henceforth we drop the corresponding symbol for simplicity. The band pairing interaction $\Gamma'_{\nu\nu'}$ is related to the corresponding orbital one as $ \Gamma'_{\nu\nu'}({\bf k,q})=\sum_{\alpha\beta\gamma\delta} \Gamma_{\alpha\beta}^{\gamma\delta}({\bf q})\phi^{\nu\dagger}_{\alpha}({\bf k})\phi^{\nu\dagger}_{\beta}(-{\bf k})\phi^{\nu'}_{\gamma}({\bf -k-q})\phi^{\nu'}_{\delta} ({\bf k+q})$. We define the SC gap in the $\nu^{\rm th}$-band as
\begin{eqnarray}
\Delta_{\nu}({\bf k})= -\frac{1}{\Omega_{\rm BZ}}\sum_{\nu',{\bf q}}\Gamma'_{\nu\nu'}({\bf k},{\bf q})\left\langle \gamma_{\nu'\sigma'}({\bf -k-q})\gamma_{\nu'\sigma}({\bf k+q})\right\rangle,
\label{SC1}
\end{eqnarray}
where the expectation value is taken over the BCS ground state. In the limit $T \rightarrow 0$ we have $\left\langle \gamma_{\nu\sigma}({\bf -k})\gamma_{\nu\sigma}({\bf k})\right\rangle \rightarrow \lambda \Delta_{\nu}({\bf k})$, with $\lambda$ as the SC coupling constant. Substituting this in Eq.~\eqref{SC1}, we get
\begin{eqnarray}
\Delta_{\nu}({\bf k})= -\lambda\frac{1}{\Omega_{\rm BZ}}\sum_{\nu',{\bf q}}\Gamma'_{\nu\nu'}({\bf k,q})\Delta_{\nu'}({\bf k+q}).
\label{SC2}
\end{eqnarray}
This is an eigenvalue equation of the pairing potential $\Gamma'_{\nu\nu'}({\bf q}={\bf k}-{\bf k^{\prime}})$ with eigenvalue $\lambda$ and eigenfunction $\Delta_{\nu}({\bf k})$. The ${\bf k}$-dependence of $\Delta_{\nu}({\bf k})$ dictates the pairing symmetry for a given eigenvalue. There are many solutions (as many as the ${\bf k}$-grid), however, we consider the highest eigenvalue since this pairing symmetry can be shown to have the lowest free energy value in the SC state.\cite{ScalapinoRMP} 

The spin-fluctuation mediated pairing potential $\Gamma'_{\nu\nu'}({\bf q})>0$, i.e. repulsive. Since we consider the highest {\it positive} eigenvalue $\lambda$, such a solution demands that the SC gap function changes sign as ${\rm sgn}\left[\Delta_{\nu}({\bf k})\right]=-{\rm sgn}\left[\Delta_{\nu'}({\bf k+q})\right]$ for those ${\bf q}$ values where $\Gamma'_{\nu\nu'}({\bf q})$ has strong contributions. As discussed in the previous section, in the weak-coupling region, $\Gamma'_{\nu\nu'}({\bf q})$ has strong peaks at the FS nesting wavevectors ${\bf Q}$. In cuprates, ${\bf Q}=(\pi,\pi)$, giving the $d$-wave symmetry to have the leading eigenvalue. In the next section, we study how the nesting feature and corresponding leading pairing symmetry solution is modified when the chain state hybridizes with the plane states.

The limitations of the weak-coupling RPA method in predicting the pairing state should be mentioned. The weak-coupling approach is more reliable at optimal doping region, as is done here, where the interaction is presumably weakened due to screening. In this limit, the other Feynman diagrams as well as vertex corrections give higher order corrections in $\mathcal{O}(U^2)$, and are less important. In addition, in the present method the pairing terms are computed over the noninteracting ground state, and no retardation effect is included. Typically, prior calculations, self-consistently including the retardation effect, obtained the same $d$-wave pairing symmetry as the weak-coupling theory predicts in cuprates.\cite{TDahm,TMaier,DManske} This agreement justifies that the salient pairing symmetry and doping dependence of the pairing eigenvalues are qualitatively reproduced within the weak-coupling theory, and the results only differ at the level of the pairing amplitudes when retardation effects are included.

\section{Results}\label{results}

\subsection{Electronic structure}
We start with the discussion of the electronic structure and FS topologies for various representative cases in Fig.~\ref{fig:orb_wgt}. For most discussions in this section, we focus on near-optimal doping region of $n_p$ = 0.82 ($x_p\approx 0.18$, $\mu_{p}=-0.35$ eV) for the plane state, and vary chain state filling factor $n_c=(0.95, 0.53, 0.15)$, corresponding chemical potential for chain states are ($\mu_{c}= -0.1, -0.9, -1.29$ eV), Figs.~\ref{fig:orb_wgt}(a)-~\ref{fig:orb_wgt}(c), respectively. The topology of the chain band allows it to accommodate electron-like FS in all cases. For the deeply electron-doped region, it forms open-orbit FS as shown in Figs.~\ref{fig:orb_wgt}(d) and ~\ref{fig:orb_wgt}(e). When the chain band becomes nearly empty, see Figs.~\ref{fig:orb_wgt}(c) and ~\ref{fig:orb_wgt}(f), the corresponding FS forms nearly closed electron-like FS [due to finite second-nearest neighbor chain-chain hopping $t_{cx}\ne 0$ along the $a$-direction]. In the intermediate filling factors ($n_c=0.53$), the FS matches those of the DFT results\cite{DFTYBCO} and ARPES data\cite{damascelli2008,damascelli2010,ARPES_borisenko} in the single crystal of YBCO$_{6+x}$ samples [e.g., Figs.~\ref{fig:orb_wgt}(e)]. 

The previously unexplored region of large filling factor $n_c$ in Figs.~\ref{fig:orb_wgt}(d) is of our prime interest here, because here we obtain an $f$-wave pairing solution, as discussed below. In this region, we find that the HHS lies above the BZ diagonal direction. In this case, we will show below that the FS nesting wavevector between the two chain FSs becomes comparable to that of the plane state and thus intervenes the overall FS nesting driven pairing potential, and hence the pairing symmetry is altered.

\subsection{Evolution of FS nesting  with chain doping}

\begin{figure*}
\includegraphics[width=0.8\textwidth]{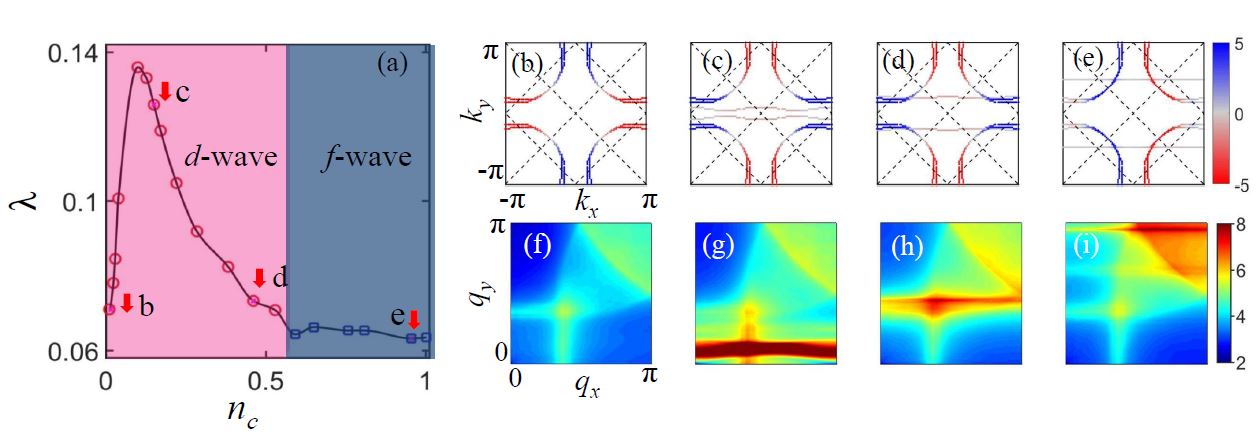}
\caption{ (a) We plot the leading SC eigenvalue (coupling constant) as a function of chain state doping. Blue square and red circles denote $f$- and $d$-wave symmetries, respectively, as the leading pairing instability. Light and dark shadings denote doping regions with $d$ and $f$- wave pairing symmetries, respectively. (b)-(e) Computed pairing eigenfunction $\Delta({\bf k})$ for the leading eigenvalue, plotted on the corresponding FSs, for four representative values of $n_c$. Here the red to blue colormap denotes the negative to positive sign of $\Delta({\bf k})$. (f)-(i) Corresponding RPA spin susceptibilities (traced over all three intra- and interlayer components) [${\rm Tr}(\tilde{\chi}_s)$] for the same cases as shown in the corresponding upper panels. All plots are shown in the same color scale for easy comparison. Here we used $\mu_{p}= -0.35$ eV and the corresponding plane state doping is $x_p\approx0.18$. 
}
\label{fig:susceptibility_lambda_comp}.
\end{figure*}

Next, we discuss the FS nesting profile as a function of chain state filling $n_c$  while keeping the plane doping fixed at $n_p\approx 0.82$, in Figs.~\ref{fig:susceptibility_lambda_comp}(f)-~\ref{fig:susceptibility_lambda_comp}(i). Here we mainly focus on the RPA spin susceptibility plotted as a function of $(q_x,q_y)$, since it contributes most to the pairing interaction. Throughout the calculation, we fix Coulomb interactions as intraband $U_{p,c}$ = 0.7, 0.6 eV, and inter-band $V_{p,c}$ = 0.5, 0.5 eV (we also explore the $U$, $V$ dependence of the results below in which the conclusions remain intact). It is easy to identify that the nearly horizontal part in the $\chi_s({\bf q})$ plot stems from the intrachain FS nesting, while the rest of the features are dominated by plane FS nestings. Of course, both nestings are affected by each other. Especially, it is worthwhile mentioning that in the case of no chain FS in Fig.~\ref{fig:susceptibility_lambda_comp}(b), the corresponding plane state nesting profile continues to break the $C_{4}$ rotational symmetry. This occurs due to plane-chain hopping $t_{cp}$ as well as their interaction $V_{c}$. A detailed layer decomposed spin susceptibility profile is given in Appendix~\ref{App:susceptibility}. 

Let us define the chain state FS nesting wavevector as ${\bf Q}_c \sim ({\rm all}~q_x, Q_{cy})$. For the plane state, the FS nesting wavevector of present interest is the one near the $(\pi,\pi)$ point, but it is incommensurate at finite dopings in all hole-doped cuprates. We denote it by ${\bf Q}^{(1)}_p\sim (\pi,Q_{py})$ and ${\bf Q}^{(2)}_p\sim(Q_{px},\pi)$. For other $C_{4}$ invariant cuprates, $Q_{px}=Q_{py}$, but it is not the case in YBCO due to coupling with the chain state. We find that in the regions of high chain state filling factor ($n_c$) $-$ when the chain FS is large and the HHS lies above the BZ diagonal $-$  $Q_{cy}\sim Q_{py}$, see Fig.~\ref{fig:susceptibility_lambda_comp}(i). This makes the total spin susceptibility to possess a dominant nesting strength at $Q_{py}$ compared to that at $Q_{px}$. As a result of the effective $C_{4}$ symmetry breaking in the spin susceptibility, and hence in the pairing interaction, the pairing eigenfunction $\Delta({\bf k})$ also acquires a symmetry which lacks this symmetry. This gives the $f$-wave symmetry.

With decrease of the chain state occupancy, the chain FS nesting wavevector becomes smaller than the plane state nesting, i.e., $Q_{cy}< Q_{py}$, and thus their contributions become decoupled. In such a case, we find that the pairing symmetry will be essentially dictated by the plane FS nesting, which gives an $d$-wave pairing. For a fixed plane layer filling factor $n_p$, the transition from the $f$-wave to $d$-wave solution occurs very much when the $Q_{cy}$ becomes smaller than $Q_{py}$. On the other hand, for $Q_{cy}\ge Q_{py}$, we find that the $f$-wave solution always dominate the $d$-wave solution. 

In the intermediate chain state occupancy when the chain FS and plane FS's van-Hove singularity merge, see Fig.~\ref{fig:susceptibility_lambda_comp}(h), the wavevector $Q_{cy}$ merges with the charge order wavevector of the plane state. This can promote a stronger and uniaxial charge ordering strength.\cite{COXray} Here, we do not investigate further the charge order state, and returns back to the pairing solution at the spin-fluctuation wavevector henceforth.

The chain band bottom is almost flat in the Cu-O bond direction. So, when the chain band becomes nearly empty, and the flat band reaches the Fermi level, its high density of states have useful ramification, see Fig.~\ref{fig:susceptibility_lambda_comp}(c). In this case, $Q_{cy}\rightarrow 0$, giving an almost massless, unidirectional paramagnon mode in the chain state, see Fig.~\ref{fig:susceptibility_lambda_comp}(g). As a result, the overall carrier concentration at the Fermi level is drastically enhanced. This enhancement optimizes the SC pairing strength as a function of chain state doping, as also obtained in the numerical result (to be discussed below). However, such a massless paramagnon mode dose not directly contribute to the unconventional pairing mechanism outlined in Sec.~\ref{theory}. For the pairing solution, the antiferromagnetic wavevector in the plane state is important, and hence we obtain an $d$-wave solution, with only a strong enhancement of the pairing strength added by large density of states of the chain state. 

Finally, as the chain state becomes completely empty, the overall FS topology and the nesting profile is dictated by the plane state. However, due to finite coupling to empty chain bands, the susceptibility topology continues to exhibit a slight loss of four-fold rotational invariance as shown in Fig.~\ref{fig:susceptibility_lambda_comp}(f). The pairing strength also decreases in this region.

\subsection{Superconducting properties}

We now turn to the main topic of superconductivity. For the same doping value where susceptibility results are discussed in the above section, we report the solutions of the largest pairing eigenvalue and pairing eigenfunction in Fig.~\ref{fig:susceptibility_lambda_comp}(a) and Figs.~\ref{fig:susceptibility_lambda_comp}(b)-~\ref{fig:susceptibility_lambda_comp}(e). The pairing eigenfunction is plotted on the corresponding FS in a colormap with blue to red colors denoting positive to negative sign of the pairing eigenfunction $\Delta({\bf k})$. The two pairing symmetry solutions we obtain have the ${\bf k}$-dependence form as (visualized over the BZ in Fig.~\ref{fig:order_parametrs})
\begin{subequations}
\begin{eqnarray}
f{\rm - wave:}~~~\Delta^{f}&=&\sin{k_x}(\cos{k_x}-3\cos{k_y}-2),\label{fwave}\quad\quad\\
d{\rm - wave:}~~~\Delta^d&=&\cos{k_x}-\cos{k_y}\label{dwave}.
\end{eqnarray}
\end{subequations}

Our nesting results reveal that when the chain nesting $Q_{cy}\ge Q_{py}$, the FS nesting at ${\bf Q}^{(1)}_p=(\pi,Q_{py})$ dominates over ${\bf Q}^{(2)}_p=(Q_{px},\pi)$. Hence the pairing potential and pairing eigenfunction inherits this broken $C_{4}$ symmetry. Moreover, the weak $q_x$ dependence of the $Q_{cy}$ nesting wavevector implies that more Fermi momenta $k_x$ are nested by this fixed wavevector, due to weak $k_x$ dispersion of the chain state as seen in Fig.~\ref{fig:orb_wgt}. This in-plane anisotropic nesting promotes a pairing symmetry which favors the condition: ${\rm sgn}\left[\Delta(k_x,k_y)\right]=-{\rm sgn}\left[\Delta(k_x+\pi,k_y+Q_{py})\right]$ at all $k_x$ - points. Owing to the FS topology of the plane state, such a condition is satisfied by $k_x\rightarrow -k_x$. As we reach the BZ boundary near ${\bf k}\sim (\pm\pi,0)$, the condition is reversed in such a way that the pairing symmetry further changes sign, see Fig.~\ref{fig:orb_wgt}(f). This is the reason, a purely $p$-wave solution (which flips signs for all $k_x\rightarrow -k_x$) is overturned by a higher-angular momentum solution with odd-parity. For the $f$-wave case, the pairing symmetry reverses sign for all $k_x\rightarrow -k_x$, in addition to another sign reversal between $k_y=0$, and $k_y=\pm \pi$ points [see Fig.~\ref{fig:order_parametrs}(a)]. As a result, we have an $f$-wave pairing state in this doping region of the chain state. 

In Fig.~\ref{fig:susceptibility_lambda_comp}(a) we plot the largest eigenvalue with blue square and red circles for $f$-wave and $d$-wave solutions, respectively. As anticipated, for large electron occupancy in the chain state which gives $Q_{cy}\ge Q_{py}$, we obtain an $f$-wave pairing solution. Otherwise, the pairing symmetry is the typical $d$-wave type. In addition, we also find that the value of the largest eigenvalue (pairing strength) gradually increases with decreasing chain state  filling factor $n_c$ (keeping everything else fixed). This increment is related to the competition between the spin fluctuation magnitude (directly enhancing the pairing strength), as well as the total density of states on the Fermi level. We notice that with decreasing chain state occupancy, the flat band of the chain state approaches the Fermi level, and hence enhances the carrier concentration. As the chain state moves completely above the Fermi level, the pairing strength again starts to decrease. This gives a new tunability to enhance superconductivity in YBCO cuprates by selectively reducing the chain states occupancy. In the existing experimental reports, such a selective tunability of the chain state is not directly explored, and hence the confirmation of our prediction awaits a focused experiment along this direction.\cite{Middey}

\begin{figure}
\includegraphics[width=0.5\textwidth]{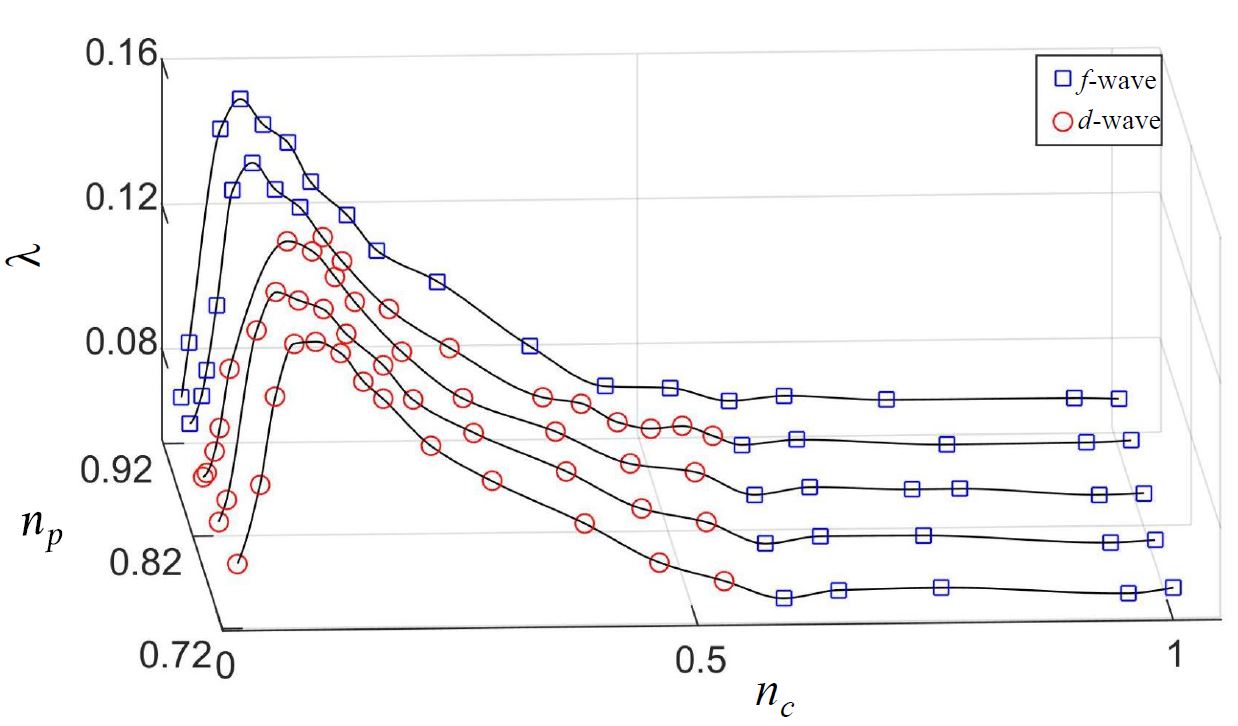}
\caption{ (a) We plot the leading pairing strength as a function of $n_c$ for several fixed values of $n_p$. In all cases, we have fixed the interaction strength and all tight-binding parameters. Blue square and red circles distinguish the leading pairing strength for $f$-wave and $d$-wave cases, respectively, and the solid line is a guide to the eye. There are prominent maxima of the pairing strength at an optimum chain doping, where the chain band bottom crosses the Fermi level. The optimum chain doping varies only weakly with the plane state doping.}
\label{fig:lambda_muc_mup} 
\end{figure}

Next we investigate the evolution of the pairing symmetry and the corresponding pairing eigenvalue $\lambda$ as a function of $n_p$ and $n_c$ in Fig.~\ref{fig:lambda_muc_mup}. Blue square and red circles distinguish between the $f$-wave and $d$-wave pairing eigenvalues, respectively, as the leading solution for a given case. We consistently find that below a critical chain filling factor $n_c$ for a fixed $n_p$, the pairing symmetry remains $d$-wave. The $d$-wave eigenvalue $\lambda$ reaches an optimum value when the chain state passes through the Fermi level. For a higher value of $n_c$, when the chain nesting vector $Q_{cy}$ becomes comparable to that of $Q_{py}$ of the plane state, the pairing symmetry changes to an $f$-wave symmetry. This condition varies for different $n_p$ values since the values of $Q_{py}$ is also doping dependent. 

Our results indicate a reentrant of the $f$-wave solution for lower hole doping on the plane state at higher values of $n_c$. In fact, with even lower hole doping, the entire $n_c$ range shows an $f$-wave solution to be dominant over the $d$-wave solution (the difference between the two eigenvalues is however very small). This occurs because the FS nesting in the plane state becomes more commensurate, tending the FS instability toward other density wave orders (such as charge density wave, spin-density wave, etc). However, the chain state nesting continues to grow and dominate over the plane state nesting. 

Caution to be taken for the results in the underdoped region. Note that our ground state in the nonSC state is a paramagnet with full FSs. The FS becomes gapped out due to charge order, pseudogap etc in the underdoped region. In fact, in the underdoped region, experiments suggest a nodeless SC gap in YBCO and other cuprates,\cite{YBCO} which presumably arises due to competition with the normal state competing orders.\cite{TSC,DasNodelessSC,Gupta} 

\begin{figure}
\includegraphics[width=0.48\textwidth]{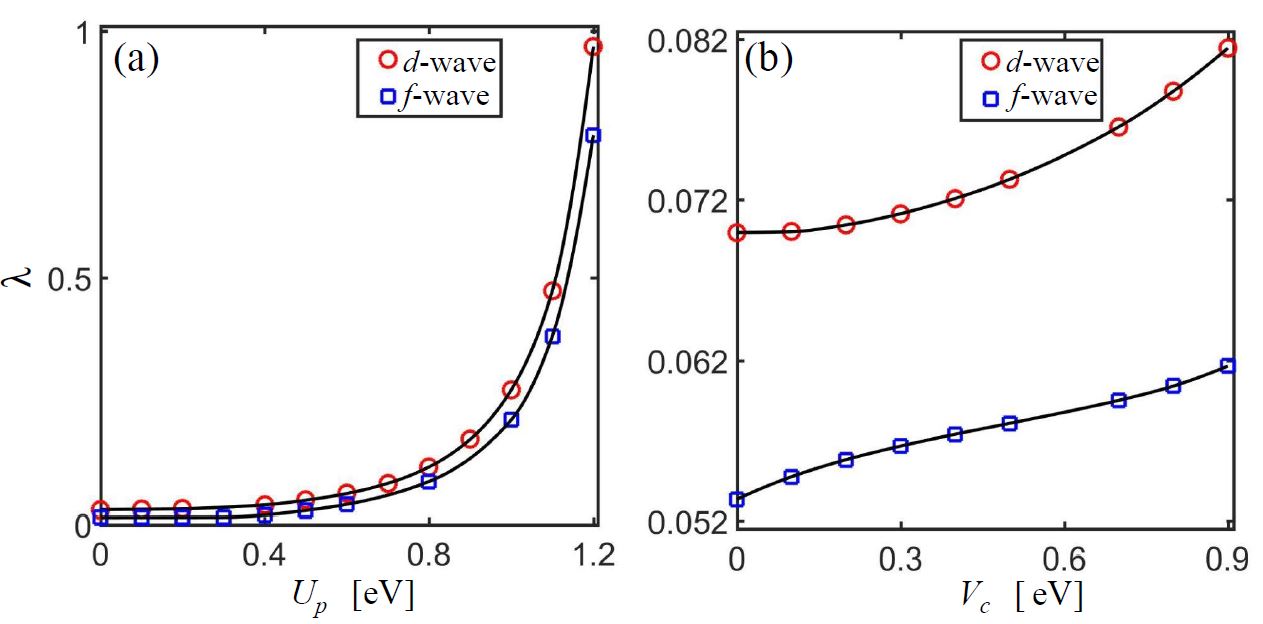}
\caption{ We plot the pairing strength $\lambda$ at two representative chain dopings, where $f$-wave and $d$-wave channels are dominant, as a function of $U_p$, and $V_c$, keeping all other parameters fixed. Here we choose $n_p$ = 0.82, $n_c$ = 0.38 ($\mu_{p}$ = -0.35, $\mu_{c}$ = -1.1 eV) for $d$-wave symmetry, and $n_p$ = 0.82, $n_c$ = 0.84 ($\mu_{p}$ = -0.35, $\mu_{c}$ = -0.3 eV) for $f$-wave symmetry for both (a)- and (b).  The solid line is a guide to the eye. The results reveal that for the doping region, where the $f$-wave eigenvalue is larger than that of the $d$-wave, this conclusion remains unchanged as a function of $U_p$ and $V_c$. For the other dopings, where $d$-wave is dominant over $f$-wave, the conclusion is also invariant for the values of $U_p$, $V_c$.}
\label{fig:lambda_vc} 
\end{figure}

\begin{figure}
\includegraphics[width=0.3\textwidth]{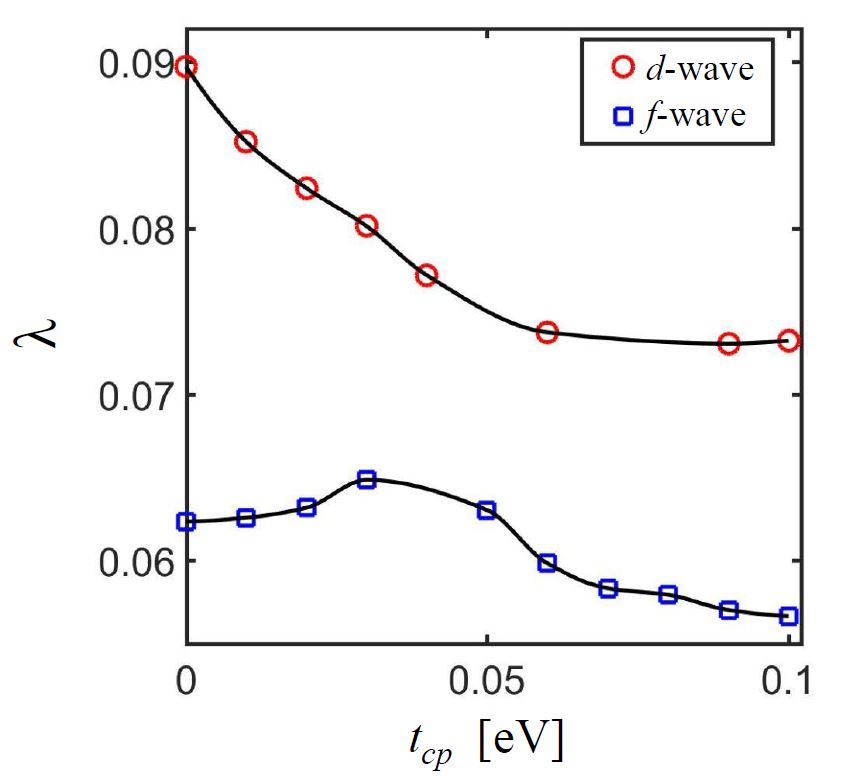}
\caption{
We plot $\lambda$ as a function of the plane-chain tunneling amplitude $t_{cp}$ on the pairing eigenvalues. Here we choose $n_p$ = 0.82, $n_c$ = 0.38 ($\mu_{p}$ = -0.35, $\mu_{c}$ = -1.1 eV) for $d$-wave symmetry, and $n_p$ = 0.82, $n_c$ = 0.94 ($\mu_{p}$ = -0.35, $\mu_{c}$ = -0.1 eV) for $f$-wave symmetry.  $U_{p/c}$ = (0.7, 0.6), $V_{p/c}$ = (0.5, 0.5) in eV. We conclude that for the doping where $d$-wave is dominant over $f$-wave, it remains so for all values of $t_{cp}$, and vice versa.}
\label{fig:lambda_tcp_new} 
\end{figure}

Finally, we address the robustness of the conclusions with respect to the interaction strength $U_p$, $V_{c}$ in Fig.~\ref{fig:lambda_vc}, as well as as a function of plane-chain hopping strength ($t_{cp}$) in Fig.~\ref{fig:lambda_tcp_new}.  We indeed find that both results are robust to the values of $t_{cp}$, $U_p$, and $V_{c}$. This  confirms that the pairing symmetry is nearly indifferent to these parameters, and is mainly determined by the FS topology and nesting profile which are dictated by filling factors. Of course, the magnitude of the pairing potential, and hence the value of the pairing eigenvalue $\lambda$ are sensitive to the energy scales of the problem which depends on $t_{cp}$, $U$, $V$.

\section{Discussions and Conclusions}\label{discuss}
Much like increasing the SC transition temperature $T_c$, obtaining varieties of unconventional pairing symmetry is an important milestone in the field of superconductivity. Especially, the odd parity pairing symmetry holds a special interests in the community in pursuit of governing triplet pairing, chiral pairing, topological superconductivity, and Majorana edge modes, etc. $f$-wave pairing symmetry is odd under  reflection along the $x$-direction and is even under reflection in the $y$-direction [see Fig.~\ref{fig:order_parametrs}(a)]. It is naturally arising in the spin-triplet channel to conform the fermionic antisymmetric wavefunction criterion, and breaks time-reversal symmetry. So far, there have been some discussions of time-reversal symmetry breaking pairing channels with $d+id$ or $s+id$ pairing channel in the spin singlet channels,\cite{Annica} or $p$-wave solutions in the spin-triplet channels\cite{TSC,DasNodelessSC,Gupta} in cuprates. However, the exploration of novel pairing channel by exploiting the chain state doping as a new tuning parameter has not been pursued before in the literature. 

Proposals of $f$-wave pairing have been put forward in heavy-fermion UPt$_3$,\cite{fHF} twisted bilayer gaphene,\cite{fTBG} monolayer MoS$_2$,\cite{fMoS2}, cold atom optical lattice,\cite{fCold} $p$-doped semiconductors,\cite{fpSC} honeycomb lattices,\cite{fspinliquid}, and other superconductors.\cite{fother} However, apart from indirect hints of such pairing symmetry in UPt$_3$,\cite{fHF} this state has not been directly realized in other families. 

The $f$-wave pairing symmetry in YBCO samples results from the competition between the chain and plane states' nesting wavevectors and strength. The plane state nesting along $(\pi,\pi)$ gives the $d$-wave symmetry. However, as the uniaxial nesting of the chain state becomes comparable in the nesting wavevector, and nesting strength to the plane state one, it breaks the $C_4$ rotational symmetry in the pairing potential. Hence the $f$-wave pairing symmetry arises. In this pairing state, the Fermi momenta change sign for all values of $k_x\rightarrow -k_x$, in addition to an additional sign reversal between $k_y=0$ and $k_y=\pm \pi$. 

Both pairing symmetries give nodal quasiparticles spectrum in the density of states, however, the gap nodes are aligned along the BZ boundary directions for the $f$-wave case, while it is aligned to the diagonal direction in the $d$-wave case. The $f$-wave pairing symmetry can also be detected by the field-angle dependence of the transport and thermodynamical quantities.\cite{BangleSC} Moreover, the anisotropy in the upper critical field in the vortex phase has unique signatures for the $f$-wave pairing as discussed in the context of UPt$_3$ superconductors.\cite{fHF}. 

As we mentioned before, the prediction of the $f$-wave pairing solution is obtained in the doping range where the carrier concentration of the chain state is substantially reduced to its intrinsic values in YBCO samples. Therefore, it is crucial to be able to dope the chain layer without altering doping concentration in the plane layers. Many organic superconductors also host a quasi-one dimensional chain state with anisotropic nesting and transport properties.\cite{Organic} Therefore, the search for an $f$-wave pairing can be easily extended to this family. 

\acknowledgements
We thank Srimanta Middey for discussions, and for showing us his preliminary experimental data on this research. The work is benefited from the computational supports from the Supercomputer Education and Research Centre (SERC) at Indian Institute of Science (IISc).

\appendix

\section{Spin susceptibility components}\label{App:susceptibility}

\begin{figure}[h]
\includegraphics[width=0.5\textwidth]{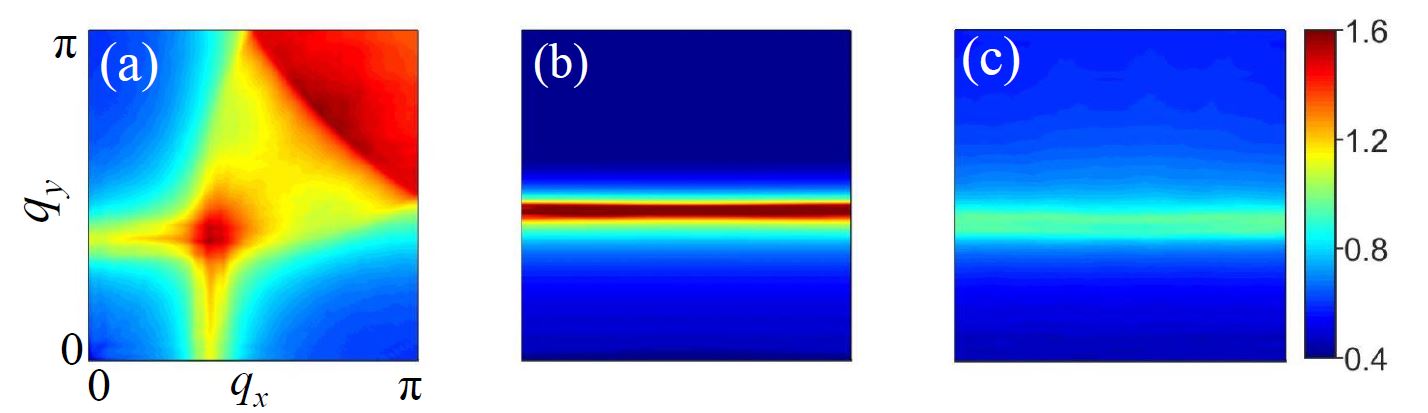}
\caption{ Computed RPA  spin susceptibility (Eq.~\eqref{spin_sus}) is split for three channels: intraplane in (a), intrachain in (b) and plane-chain in (c). Filling factors are $n_{p}=0.82$ and $n_{c}=0.65$. $U_{p/c}$ = 0.7, 0.6 eV, and $V_{p/c}$= 0.5, 0.5 eV. All plots are done in the same colorbar.}
\label{fig:chi_inter-band} 
\end{figure}

In Fig.~\ref{fig:chi_inter-band} , we separately show the contributions of the intraplane, intrachain and plane-chain susceptibilities for the spin-channels only. We notice that the FS nesting in the plane channel is very similar to the ones obtained in other cuprates without a chain state. The intrachain FS nesting is almost one dimensional with very weak anisotropy in the intensity. This is due to low $k_x$-dispersion at finite filling factor. The interlayer plane-chain FS nesting is also quasi-1D with significantly low intensity.

\begin{figure}[h]
	\includegraphics[width=0.5\textwidth]{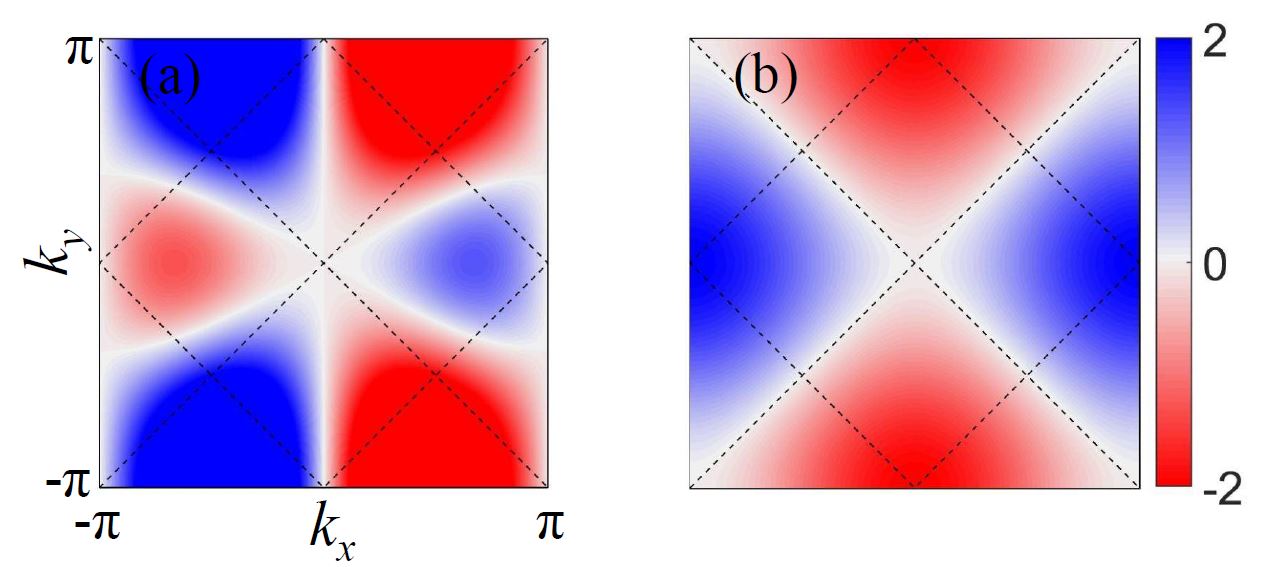}
	\caption{We visualize the ${\bf k}$-dependence of the SC pairing symmetries in (a) for $f$-wave (Eqs.~\eqref{fwave}), and in (b) for $d$-wave [Eq.~\eqref{dwave}]. The colormap of red to blue gives negative and positive signs. We did not normalize the eigenfunctions in any of the results in the main text, since normalization simply gives a constant multiplication to the eigenfunctions.}
	\label{fig:order_parametrs} 
\end{figure}

In Fig.~\ref{fig:order_parametrs}, we plot the pairing functions, Eqs.~\eqref{fwave} and \eqref{dwave}, in the 2D BZ. This plot is shown to ease the discussion of the pairing symmetry in the main paper.

\end{document}